\documentclass[sigconf]{acmart}

\usepackage{booktabs} 

\usepackage{graphicx,color,colortbl,psfrag,subfig}
\usepackage{multirow}
\usepackage{balance}

\usepackage{algorithm}
\usepackage{algpseudocode}
 
\usepackage{bigints}
\usepackage{setspace}

\usepackage{amsthm}

\usepackage{amssymb}
\usepackage{amsmath}
\usepackage{amsfonts}
\usepackage{textcomp}

\graphicspath{{}}

\newtheorem{theorem}{Theorem}
\newtheorem{definition}{Definition}

\copyrightyear{2021}
\acmYear{2021}
\setcopyright{acmcopyright}
\acmConference[KDD '21]{Proceedings of the 27th ACM SIGKDD Conference on Knowledge Discovery and Data Mining}{August 14--18, 2021}{Virtual Event, Singapore}
\acmBooktitle{Proceedings of the 27th ACM SIGKDD Conference on Knowledge Discovery and Data Mining (KDD '21), August 14--18, 2021, Virtual Event, Singapore} 
\acmPrice{15.00}
\acmDOI{10.1145/3447548.3467115}
\acmISBN{978-1-4503-8332-5/21/08}

\begin{document}

\title{Incorporating Prior Financial Domain Knowledge into Neural Networks for Implied Volatility Surface Prediction}

\author{Yu Zheng}
\affiliation{\institution{$^1$~School of Finance}}
\affiliation{\institution{Southwestern University of Finance and Economics}}
\affiliation{\institution{$^2$~Inboc Technologies Ltd.}}
\email{zhengyu@swufe.edu.cn}

\author{Yongxin Yang}
\affiliation{\institution{$^1$~School of Informatics}}
\affiliation{\institution{University of Edinburgh}}
\affiliation{\institution{$^2$~ArrayStream Technologies Ltd.}}
\email{yongxin.yang@ed.ac.uk}

\author{Bowei Chen}
\authornote{Corresponding author.}
\affiliation{\institution{Adam Smith Business School}}
\affiliation{\institution{University of Glasgow}}
\email{bowei.chen@glasgow.ac.uk}



\begin{abstract}

In this paper we develop a novel neural network model for predicting implied volatility surface.  Prior financial domain knowledge is taken into account. A new activation function that incorporates volatility smile is proposed, which is used for the hidden nodes that process the underlying asset price.  In addition, financial conditions, such as the absence of arbitrage, the boundaries and the asymptotic slope, are embedded into the loss function. This is one of the very first studies which discuss a methodological framework that incorporates prior financial domain knowledge into neural network architecture design and model training. The proposed model outperforms the benchmarked models with the option data on the S\&P 500 index over 20 years. More importantly, the domain knowledge is satisfied empirically, showing the model is consistent with the existing financial theories and conditions related to implied volatility surface. 

\end{abstract}

%
%

%



\maketitle

\section{Introduction}
\label{sec:intro}

%
%

Machine learning algorithms are essentially data-driven models which mainly focus on producing accurate predictions. They are being used for a wider array of macro and micro level prediction tasks. According to Ernst and Young~\cite{EY_2018_Fintech}, machine learning applications in finance have become one of the hottest sectors globally, with the expected direct investment growth of 63\% from 2016 to 2022. Despite COVID-19, a recent survey from the Bank of England showed that there are still a third of banks would increase their investments in the number of planned or existing machine learning and data science projects~\cite{BoE_2020}. Therefore, it is interesting for machine learning researchers and financial analysts to investigate this sector because it generates a lot of practical questions and challenges, and addressing them will result in positive economic and social consequence.

Although with excellent prediction performance, machine learning is usually used as the \lq\lq{}black box\rq\rq{} model in many financial applications. Compared with the well-developed models from mathematical finance, machine learning algorithms are less interpretable, e.g., features can be not understandable, and the learning process is not transparent or mathematically tractable. More importantly, they are not aligned with the well-developed financial theories. Therefore, many financial institutions are slow to adopt machine learning algorithms (particularly neural networks) into their major business operations. Developing interpretable machine learning models that are consistent with the existing financial markets and theories will resolve the bottleneck and will boost the applications of machine learning into finance.

In this paper, we propose a novel neural network model tailored for implied volatility surface prediction. Implied volatility is an important financial metric or indicator that captures the market's view of the likelihood of changes in a given asset price. Technically, an \emph{implied volatility} is defined as the inverse problem of option pricing, mapping from the option price of the asset in the current market to a single value~\cite{Cont_2002}. When it is plotted against the option strike price and the time to maturity, it is called the \emph{implied volatility surface}. Prior financial domain knowledge related to implied volatility surface includes: 1) the empirical evidence volatility smile; and 2) financial conditions such as the absence of arbitrage, the boundaries and the asymptotic slope. We use different ways to incorporate these domain knowledge. For the former, a new activation function that produces volatility smile is proposed, and it is used for the hidden nodes that process the underlying asset price. For the latter, financial conditions are embedded into the loss function for neural network training. In the experiments, we validate the proposed model with the option data on the S\&P 500 index over a period of 20 years. Compared with the existing studies, our experimental settings are more challenging and this requires our model to be more robust and stable in producing convincing results. Our model outperforms the widely used state-of-the-art model in finance and other benchmarked neural network models on the mean average percentage error in both training and test sets. In the meantime, the incorporated prior financial domain knowledge are met empirically.


Technology wise, our study makes a methodological contribution. We propose a framework of incorporating prior financial domain knowledge into neural network design and training. Therefore, the developed model is aligned well with the existing empirical evidence and financial theories related to implied volatility surface. This is an important step for interpretable machine learning, and we hope the framework can motivate many other investigations of machine learning applications in finance. On the other hand, from the application perspective, we develop a best-performing prediction model, and to the best of our knowledge, this is one of the very first neural networks tailored for implied volatility surface.

The rest of the paper is organised as follows. Section~\ref{sec:related_work} reviews the related literature. Section~\ref{sec:model} introduces our proposed model for predicting implied volatility surface. The used dataset, our experimental settings and results are presented in Section~\ref{sec:experiments}. Finally, we conclude the paper in Section~\ref{sec:conclusion}.

\section{Related Work}
\label{sec:related_work}

Our research in this paper touches upon two streams of literature: mathematical finance and machine learning. For the former, we introduce the basic concepts and the related studies of option pricing and volatility modelling. For the latter, we review a number of recent applications of machine learning in finance, with special focuses on option pricing and volatility modelling. 

\subsection{Option Pricing and Volatility Modelling}

In 1973, Black and Scholes~\cite{Black_1973} proposed an elegant closed-form pricing formula for the European style call options written on financial assets. Their model is simply called the \emph{Black-Scholes option pricing model}, in which an underlying financial asset price is driven by a geometric Brownian motion~\cite{Samuelson_1965} that contains a drift and a volatility, and the volatility term shows the small fluctuations of asset returns representing risk. The seminal work of Black and Scholes opened the floodgates of studying mathematical models in finance, and volatility models have soon become popular since then~\cite{Sundaresan_2000,Friz_2005}.

Volatility models in finance can be classified into two groups~\cite{Homescu2011}. The first group is called \emph{indirect methods}, in which an implied volatility is driven by another dynamic model such as local volatility models, stochastic volatility models and L\'{e}vy models~\cite{Merton76optionpricing,Heston1993,Kou2002jump,Kang_2017,Shiraya_2018}. Models in this group usually have a limited number of parameters, and the volatility term is fitted by the market data along with the asset dynamics such as the geometric Brownian motion and the mean-revision jump-diffusion process. These models exhibit mathematical elegance but are sometimes invalid empirically. Time-dependent parameters can be included but they will greatly increase computational time and optimisation difficulty in model calibration. The second group is called \emph{direct methods}, in which an implied volatility is specified explicitly. Direct methods can also be divided into two types. The first type specifies the dynamics of an implied volatility surface and assumes it evolves continuously over time~\cite{Cont_2002,Carr2010}. The second type focuses on the static representation of implied volatility surface that uses either parametric or non-parametric methods to fit an implied volatility surface and then for prediction~\cite{Kotze2013,Itkin2015,Corlay2016}. Our proposed method is a static model. In this group, the stochastic volatility inspired (SVI) model is the most commonly used method~\cite{Gatheral2004}. It models the implied volatility slice for a fixed time to maturity. Gatheral and Jacquier then further improved the SVI model with a simpler representation on the conditions for no static arbitrage, and this improved SVI model is called the \emph{surface SVI (SSVI)} model, which is the recent advance in mathematical finance and has been widely adopted by investors~\cite{Gatheral2014}. Therefore, we choose the SSVI model as one of the benchmarked models in this paper.

\begin{figure*}[t]
\centering
\includegraphics[width=0.875\linewidth]{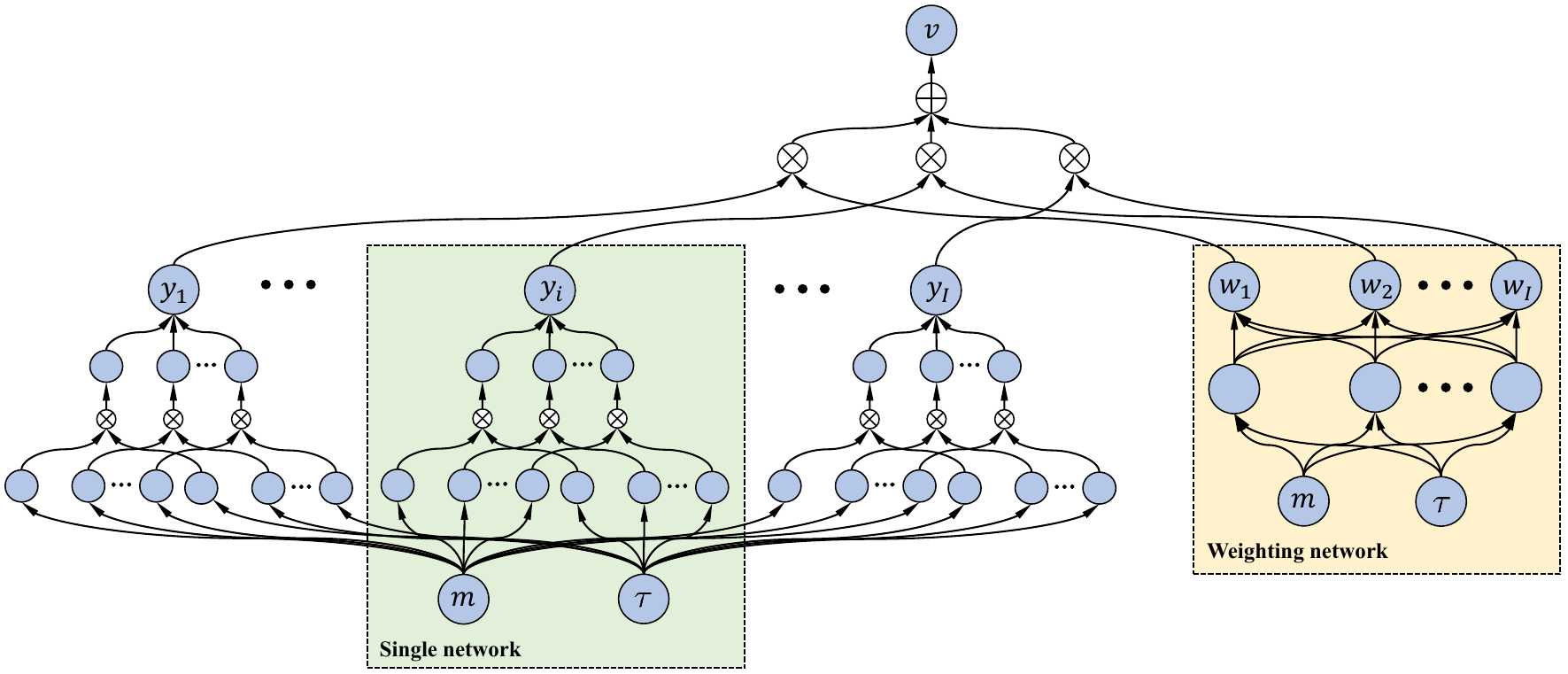}
\vspace*{5pt}
\caption{Neural network architecture design tailored to implied volatility surface. The proposed multi model consists of several single models and their weights are determined by the weighting network. Bias terms are omitted; $\otimes$ is the multiplication gate operator; $\oplus$ is the addition gate operator. }
\label{fig:neural_network_structure}
\end{figure*}

\subsection{Machine Learning Applications in Finance}

Applying machine learning to asset pricing and volatility prediction can be traced back to the late 1980\rq{}s or early 1990\rq{}s. In the early stages, single hidden layer neural networks were used to estimate option price~\cite{Malliaris_1993} and to predict the volatility of the S\&P 100 index~\cite{Malliaris_1996}. Then, various machine learning algorithms were introduced, including ensemble methods~\cite{Gavrishchaka2006,Audrino2010}, kernel machines~\cite{Coleman2013}, Gaussian process~\cite{Wu_2014}, and deep learning models such as hybrid neural networks~\cite{Kristjanpoller_2014}, gated neural networks~\cite{Yang_2017}, and recurrent neural networks~\cite{Luo_2018}. In addition to the conventional financial data, several other recent studies developed models which use verbal, vocal and social information~\cite{Rekabsaz_2017,Qin_2019,Zheng_2019}. 

It should be noted that our research in this paper focuses on predicting the implied volatility surface when the underlying asset price and the time to maturity of the corresponding option quotes are given. This kind of prediction in essence is an inverse engineering of option pricing and it is not time series forecasting. Therefore, our scenario is different to the above mentioned studies. Our neural network architecture design is inspired by the work of~\cite{Yang_2017} but has three significant differences. First, our model aims to predict the implied volatility rather than option price. Second, we design a new activation function that can incorporate volatility smile. Third, we embed the conditions related to implied volatility into neural network training.

\section{Model}
\label{sec:model}

In this section, we firstly introduce the preliminaries of implied volatility surface and lays down the mathematical settings. We then discuss our deep neural network architecture design and further explain how do we incorporate the prior financial domain knowledge into neural network.

\subsection{Prior Financial Domain Knowledge}
\label{sec:setup}

In mathematical finance, the spot price of an asset is usually modelled as a stochastic process $(S_t)_{t\geq 0}$ that is defined on a filtered probability space $(\Omega,\mathcal{F}, ({\mathcal{F}}_t)_{t\geq 0}, \mathbb{P})$, where $t$ is the time index, $\Omega$ is the sample space, $\mathcal{F}$ is the sigma-field, $({\mathcal{F}}_t)_{t\geq 0}$ is the filtration, and $\mathbb{P}$ is the probability space. The financial market is assumed to be arbitrage-free and a financial product's time to maturity (i.e., the time remaining until a financial contract expires) is always finite. 

As mentioned previously, implied volatility is the inverse engineering of option pricing. It can be obtained by inverting the Black–Scholes option pricing model~\cite{Black_1973}, in which one needs to determine the constant interest rate and dividends from the market data. To avoid dealing with them, the forward measure can be used instead. Let $(F_{t,T})_{t\geq 0}$ be the forward price of the asset with maturity date $T$, where $0 \leq t \leq T$. Then $$F_{t,T} = \frac{S_t}{B(t, T)},$$ 
where $B(t, T)$ is the price of a zero-coupon bond at time $t$ which will pay one unit at time $T$. The absence of arbitrage ensures there exists an equivalent martingale measure in which $(F_{t,T})_{t\geq 0}$ is a martingale~\cite{Cont_2002}. In probability theory, a \emph{martingale} is a stochastic process for which, at a particular time, the conditional expectation of the next value in the process is equal to the present value and regardless of all the previous values. So the log forward moneyness $m$ can be defined and used as the underlying, where $m = \log\{K/F_{t,T}\}$ and $K$ is the strike price. 

Another important variable is the time to maturity, which can be defined as 
$$\tau = \frac{T-t}{A}, $$ 
where $A$ is the annualization factor. Therefore, in our mathematical setting, the implied volatility $v$ can be written as a function of the log forward moneyness $m$ and the time to maturity $\tau$.

\begin{theorem}
\label{thm:conditions}
Let $d_{\pm}(m,\tau) = -\frac{m}{\sqrt{\tau}v(m,\tau)} \pm \frac{1}{2}\sqrt{\tau}v(m,\tau)$, $n(\cdot)$ denote the density function of a standard normal distribution, and $N(\cdot)$ denote its cumulative function. The following conditions are required to be met for the implied volatility $v$: 
\begin{enumerate}
\item[1)] \emph{(Positivity)} For $(m,\tau) \in \mathbb{R} \times \mathbb{R}^+$, $v(m,\tau)>0$.
\item[2)] \emph{(Twice Differentiation)} For $\tau>0$, $m \rightarrow v(m,\tau)$ is twice differentiable on $\mathbb{R}$.
\item[3)] \emph{(Monotonicity)} For $m \in \mathbb{R}$, $\tau \rightarrow \sqrt{\tau} v(m,\tau)$ is increasing on $\mathbb{R}^+$, then 
$$v(m,\tau)+2 \tau \partial_{\tau} v(m,\tau) \geq 0.$$
\item[4)] \emph{(Absence of Butterfly Arbitrage)} For $(m,\tau) \in \mathbb{R} \times \mathbb{R}^+$, 
\begin{align*}
\bigg[1-\frac{m \partial_{m} v(m,\tau)}{v(m,\tau)} \bigg]^2 - \frac{1}{4}\Big[v(m,\tau) \tau \partial_{m} v(m,\tau)\Big]^2 & \\
+ \ \tau v(m,\tau) \partial_{mm} v(m,\tau) & \geq 0.
\end{align*}
\item[5)] \emph{(Limiting Behaviour)} If $\tau>0$, then 
\[
\lim_{m\rightarrow + \infty} d_{+}(m,\tau) = -\infty.
\]
\item[6)] \emph{(Right Boundary)} If $m \geq 0$, then
\[
N(d_{-}(m,\tau))- \sqrt{\tau}\partial_{m} v(m,\tau)n(d_{-}(m,\tau)) \geq 0.
\]
\item[7)] \emph{(Left Boundary)} If $m < 0$, then
\[
N(-d_{-}(m,\tau))+ \sqrt{\tau}\partial_{m} v(m,\tau)n(d_{-}(m,\tau)) \geq 0.
\]
\item[8)] \emph{(Asymptotic Slope)} If $\tau>0$, then 
$$2 |m| - v^2(m,\tau)\tau>0.$$ 
\end{enumerate}
\end{theorem}

Simply, Theorem~\ref{thm:conditions} conditions 1-5 ensure the absence of arbitrage~\cite{Gulisashvili_2012}; conditions 6-7 specify the boundaries~\cite{Carr_2007}; and condition 8 is the asymptotic slope~\cite{Lee_2004}. 

In addition to Theorem~\ref{thm:conditions}, implied volatility has an important empirical evidence (or stylised fact) called the \emph{volatility smile} -- for a given time to maturity, when the implied volatility is plotted against the strike price, it creates a line that slopes upward on either end, looking like a \lq\lq{}smile\rq\rq{}~\cite{Cont_2002}. In the following discussion, we will present our neural network model by incorporating these mentioned financial conditions and empirical evidence related to implied volatility surface.

\subsection{Deep Neural Network Architecture Design}
\label{sec:design}

Figure~\ref{fig:neural_network_structure} presents a schematic view of our neural network architecture. The model input includes the log forward moneyness $m$ and the time to maturity $\tau$, and the model output is the implied volatility $v$. The proposed neural network is constructed from sub-networks with two types of architectural structures: 1) several simple networks (called \emph{single networks}) which predict the implied volatility separately; and 2) a weighting network which determines the \lq\lq{}votings\rq\rq{} of the predicted implied volatilities towards the final prediction. Similar to~\cite{Yang_2017}, multiplication and addition gate operators are used to process and merge information related to the log forward moneyness $m$, the time to maturity $\tau$ and single networks.

\begin{definition}[\bf Smile Function]
\label{def:smile_function}
For any $z \in \mathbb{R}$,
\begin{equation}
\label{eq:smile_function}
\phi(z) = \bigg[ z \tanh(z + \frac{1}{2}) + \tanh(-\frac{1}{2} z + \epsilon) \bigg]^{1/2} \hspace*{-10pt}, \hspace*{10pt} z \in \mathbb{R},
\end{equation}
where $\tanh(\cdot)$ is the hyperbolic tangent function and $\epsilon$ is a small value to ensure numerical stability. 
\end{definition}

\begin{figure}[t]
\centering
\includegraphics[width=0.9\linewidth]{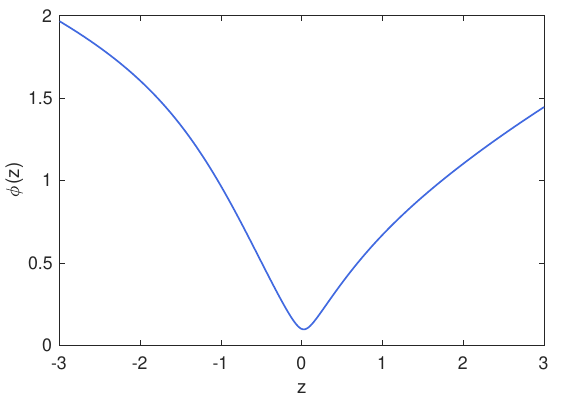}
\caption{Smile function $\phi(\cdot)$ where $\epsilon=0.01$.}
\label{fig:smile_function}
\end{figure}

As shown in Figure~\ref{fig:smile_function}, the defined smile function exhibits a skew pattern like volatility smile. We apply the smile function $\phi(\cdot)$ to the nodes that correspond to $m$ and the sigmoid function $\psi(\cdot)$ for the nodes that correspond to $\tau$. It is not difficult to see that the positivity and twice differentiation conditions in Theorem~\ref{thm:conditions} are met and the limiting behaviour condition can be proven theoretically by inverting the Black-Scholes option pricing model~\cite{Gulisashvili_2012}.

Our proposed neural network can be expressed as follows:
\begin{align}
\hat{v} 
= & \ 
\sum_{i=1}^{I} y_i w_i, \label{eq:multi}\\
y_i 
= & \
\sum_{j=1}^{J}  
\phi(m \bar{w}_{j}^{(i)} + \bar{b}_j^{(i)})
\psi(\tau \tilde{w}_{j}^{(i)} + \tilde{b}_j^{(i)})
e^{\hat{w}_{j}^{(i)}} + e^{\hat{b}^{(i)}} \hspace*{-5pt},\\
w_i 
= & \ 
\frac{\exp\bigg\{\sum_{k=1}^{K}\psi(m\dot{w}_{1,k} + \tau\dot{w}_{2,k} + \dot{b}_k) \ddot{w}_{k,i} + \ddot{b}_i \bigg\}}{ \sum_{i=1}^{I} \exp \bigg\{\sum_{k=1}^{K}\psi(m\dot{w}_{1,k} + \tau\dot{w}_{2,k} + \dot{b}_k) \ddot{w}_{k,i} + \ddot{b}_i \bigg\}
}, 
\end{align}
where $I$ is the number of single networks, $J$ is the number of layers in each single network, and $K$ is the number of layers in the weighting network. For the $i$th single network, $y_i$ is its prediction output. To ensure it is non-negative, we consider the exponential form of the weight $e^{\hat{w}_j^{(i)}}$ and the bias $e^{\hat{b}^{(i)}}$. For the hidden layers, $\bar{w}_j$ and $\bar{b}_j$ are the weight and the bias of the $j$th hidden node that corresponds to $m$, and $\tilde{w}_j$ and $\tilde{b}_j$ are the weight and the bias of the $j$th hidden node that corresponds to $\tau$. Therefore, there is a total of $5J+1$ parameter values for each single network. The weighting network predicts the weights of single networks towards the final prediction, where $w_i$ is its prediction for the $i$th single network, $\dot{w}_{\cdot,k}$ is the weight of the $k$th hidden node, $\dot{b}_k$ is the bias of the $k$th hidden node, $\ddot{w}_{\cdot,i}$ is the weight of the $i$th single network, and $\ddot{b}$ is the bias of the $i$th single network. Since the dimensions of $\dot{w}$, $\dot{b}$, $\ddot{w}$, $\ddot{b}$ are $2\times K$, $K \times 1$, $K\times I$, and $I \times 1$, respectively, the total number of parameter values in the our model is $(5J+K+2)I + 3K$.

\subsection{Embedding Constraints into Optimisation}
\label{sec:optimisation}

Training the designed neural network solves an optimisation problem that finds for parameters which result in a minimum loss when evaluating the samples in the training data. We define a loss function tailored to implied volatility surface by embedding the related conditions from mathematical finance:
\begin{align}
\ell 
= & \ 
\ell_0 + \gamma\ell_1 + \delta\ell_2 + \eta\ell_3 + \rho\ell_4 + \omega\ell_5,
\label{eq:loss_function_all}
\end{align}
where $\ell_0$ represents the data loss, $\ell_1, \ldots, \ell_4$ are the loss functions that incorporate financial conditions discussed previously, and $\ell_5$ is the regularization term to avoid over-fitting. 

The data loss $\ell_0$ is defined as a joint loss from the mean squared log error (MSLE) and the mean squared percentage error (MSPE):
\begin{align}
\ell_0 
= & \ 
\frac{\alpha}{N} \sum_{n=1}^{N} \big(\log v_n - \log \hat{v}_n \big)^2 \hspace*{-5pt} + \frac{\beta}{N} \sum_{n=1}^{N}\Big(\frac{v_n - \hat{v}_n}{v_n}\Big)^2 \hspace*{-5pt},
\label{eq:ell_0}
\end{align}
where $N$ is the total number of the training samples, $v_n$ is the ground truth implied volatility for the $n$th sample, $\hat{v}_n$ is the predicted implied volatility for the $n$th sample,  $\alpha$ and $\beta$ are hyperparameters that control the weights of MSLE and MSPE, respectively. We use the joint data loss here because it is efficient in dealing with sensitive data or high-dimensional feature spaces~\citep{Goodfellow_2016}.

The monotonicity condition is specified by $\ell_1$, defined by
\begin{align}
\ell_1 
= & \ 
\sum_{p=1}^{P}\sum_{q=1}^{Q} \max\{0, -a(m_p,\tau_q)\},
\label{eq:ell_1}
\end{align}
where $P$ and $Q$ are the number of samples, and $a(m,\tau) = v(m,\tau)+2 \tau \partial_{\tau} v(m,\tau)$. The objective of $\ell_1$ is to push $a(m,\tau)$ to be non-negative. This can be achieved by randomly sampling $P$ unique values from the domain of $m$ and $Q$ unique values from the domain of $\tau$. Penalty is added by $\ell_1$ if $a(m,\tau)$ is negative for the sampled $(m,\tau)$ pairs. 

The absence of a butterfly arbitrage condition is specified by $\ell_2$, defined as follows
\begin{align}
\ell_2 
= & \ 
\sum_{p=1}^{P}\sum_{q=1}^{Q} \max\{0, -b(m_p,\tau_q)\}, 
\label{eq:ell_2}
\end{align}
where
\begin{align*}
b(m,\tau) = & \ (1-\frac{m \partial_{m} v(m,\tau)}{v(m,\tau)})^2 - \frac{(v(m,\tau) \tau \partial_{m} v(m,\tau))^2}{4} \\ 
& \ + \tau v(m,\tau) \partial_{mm} v(m,\tau).
\end{align*}
The objective of $\ell_2$ is to push $b(m,\tau)$ to be non-negative. This can be achieved using the same way as $\ell_1$, by randomly sampling $P$ unique values from the domain of $m$ and $Q$ unique values from the domain of $\tau$. 

The left and the right boundary conditions are specified by $\ell_3$, defined as follows
\begin{align}
\ell_3 
= & \ 
\sum_{p_1=1}^{P_1}\sum_{q=1}^{Q} \max\{0, -c_1(m_{p_1},\tau_q)\} \nonumber \\
& \ 
+ \sum_{p_2=1}^{P_2}\sum_{q=1}^{Q} \max\{0, -c_2(m_{p_2},\tau_q)\}.
\label{eq:ell_3}
\end{align}
where 
\begin{align*}
c_1(m,\tau) = & \ N(d_{-}(m,\tau))- \sqrt{\tau}\partial_{m} v(m,\tau)n(d_{-}(m,\tau)), \\
c_2(m,\tau) = & \ N(-d_{-}(m,\tau))+ \sqrt{\tau}\partial_{m}  v(m,\tau)n(d_{-}(m,\tau)).
\end{align*}
The objective is to push both functions to be non-negative. To achieve this, we sample $P_1$ unique non-negative values from the domain of $m$, $P_2$ unique negative values from the domain of $m$ and $Q$ unique values from the domain of $\tau$.

The asymptotic condition is specified by $\ell_4$, defined by
\begin{align}
\ell_4 
= & \ 
\sum_{p=1}^{P}\sum_{q=1}^{Q} \max\{0, -(g(m_p,\tau_q)-\epsilon)\},
\label{eq:ell_4}
\end{align}
where 
$$g(m,\tau) = 2|m| - v^2(m,\tau)\tau,$$ 
and $\epsilon$ is a very small value. The aim of $\ell_4$ is to ensure $g(m,\tau)$ is positive. Therefore, similar to $\ell_1, \ell_2, \ell_3$, we sample $P$ unique values from the domain of $m$ and $Q$ unique values from the domain of $\tau$. It is worth mentioning that the values of $m$ and $\tau$ in $\ell_1, \cdots, \ell_4$ can also be sampled from the training data. However, the trained neural network may fail to meet those conditions when the given values of $m$ and $\tau$ for prediction are out of the scope of the training data. If the training data have limited observations of input variables, creating synthetic data by sampling values from their domains is an effective way to train the model with good generalization capabilities~\citep{Choe_2017,Liu_2017}

The regularization term $\ell_5$ is defined by
\begin{align}
\ell_5 
= & \ 
\sum_{i=1}^{I}||\bar{w}^{(i)}||_F^2 + \sum_{i=1}^{I}||\tilde{w}^{(i)}||_F^2  + \sum_{i=1}^{I}||\hat{w}^{(i)}||_F^2 \nonumber \\
& \ + ||\dot{w}||_F^2 + ||\ddot{w}||_F^2,
\label{eq:ell_5}
\end{align}
where $||\cdot||_F^2$ is the squared Frobenius norms.

\section{Experiments} 
\label{sec:experiments}

The used datasets are firstly introduced. We then provide the details of experimental design, including examined models and their training settings. The experimental results are finally presented and discussed. 

\subsection{Data} 
\label{sec:data}

\begin{figure}[t]
\centering
\includegraphics[width=1\linewidth]{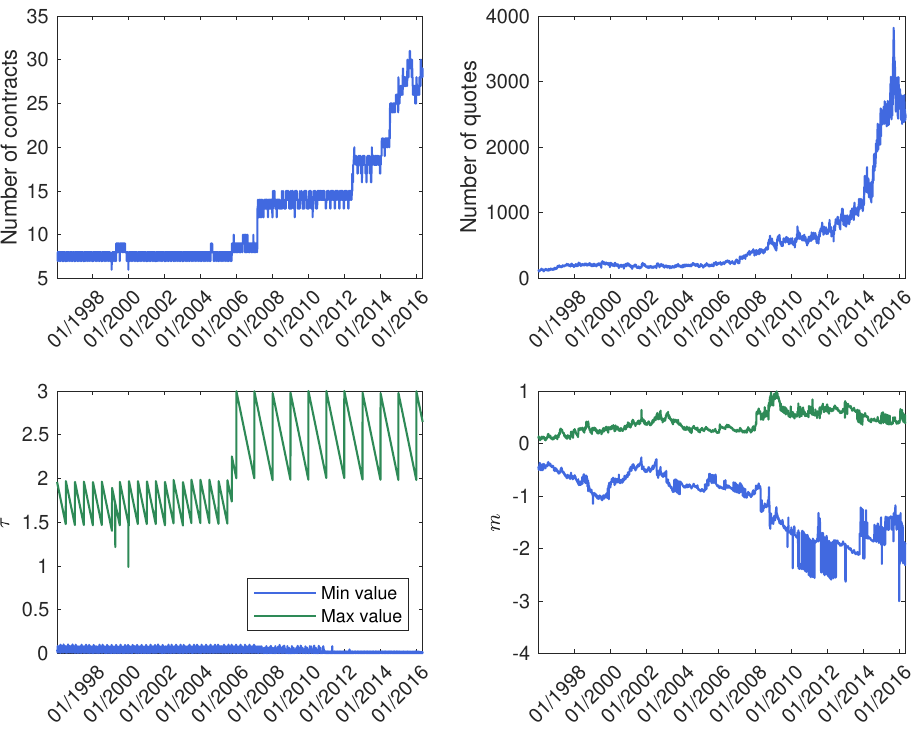}
\caption{Time series plots of cleaned option contracts and their quotes from 04/01/1996 to 29/04/2016}
\label{fig:data_summary}
\end{figure}

We use the option and the zero-coupon yield curve data from OptionMetrics, and the Overnight Index Swap (OIS) data from Bloomberg.\footnote{OptionMetrics is a leading provider of historical implied volatility, greeks, and option pricing data for financial markets and Bloomberg is a premier financial services company.} Our option data is for the S\&P 500 index. It is one of the most commonly followed stock market indices which measures the stock performance of 500 large companies listed on stock exchanges in the United States. The option data contains a total of 5,116 trading days, covering the period from 04/01/1996 to 29/04/2016. The zero-coupon yield curve shows the relationship between the level of the interest rate (representing the cost of borrowing) and the time to maturity.  It is constructed based on the London Inter-Bank Offered Rate (LIBOR). However, after the 2018 financial crisis, the LIBOR-based zero curve is not risk-free~\cite{Ametrano2013}. Therefore, adjustments are performed for the period after 01/01/2008. Specifically, we extract the OIS data from Bloomberg and bootstrap the zero rate curve. In addition, we use the cubic spline to interpolate the risk-free rates in order to match the option maturity, and compute the forward price using the put-call parity~\cite{Bilson2015}.

The option data is further processed. Option quotes which are less than 3/8 are excluded because they are close to the tick size and can be misleading. The bid-ask mid-point price is calculated as a proxy for the closing price. The in-the-money option quotes are excluded because of the small transaction volume~\cite{Bliss2005}. The existing studies usually do not analyse option contracts with the time to maturity of less than 7 days~\cite{Andersen2017}. However, as these options are getting popular recently (e.g., weekly index options), we here analyse option contracts with a short time to maturity and only exclude the contracts with the maturity of less than 2 days. Analysing options with a short maturity is challenging because this requires our model with high robustness and stability. As shown in Figure~\ref{fig:data_summary}, our prepared data finally contains 63,338 option contracts with 2,986,754 valid quotes. The quotes are then used to compute the ground truth implied volatility values by inverting the Black–Scholes option pricing formula.

\begin{table}[t]
\centering
\begin{tabular}{c|p{2.65in}}
\hline
Model 	& \multicolumn{1}{c}{Description}\\ 
\hline
SSVI 	& \cite{Gatheral2014}\\
\hline 
Multi 	& The proposed model specified in Eqs.~(\ref{eq:multi})-(\ref{eq:ell_5}).\\ 
\hline
Multi$^\dag$ & The Multi model trained without embedding $\ell_1, \ell_2, \ell_3, \ell_4$.\\ 
\hline
Single 	& The single network model so there is no weighting network, and $||\dot{w}||_F^2$ and $||\ddot{w}||_F^2$ are not included in the regularization term $\ell_5$ for the model training\\ 
\hline
Single$^\dag$ 	& The Single model trained without embedding $\ell_1, \ell_2, \ell_3, \ell_4$.\\ 
\hline
Vanilla 	&   The neural network model with the simplest architecture -- it has a single hidden layer which only uses the sigmoid activation function and the model's output is censored to be non-negative. \\ 
\hline
Vanilla$^\dag$ 	& The vanilla model trained without embedding $\ell_1, \ell_2, \ell_3, \ell_4$.\\ 
\hline
\end{tabular}
\vspace*{10pt}
\caption{Summary of the examined models.}
\label{tab:models}

\vspace*{10pt}

\centering
\begin{tabular}{c|c|c|c|c|c|c|c|c|c|c}
\hline
\multirow{2}{*}{Model}	 & \multicolumn{10}{c}{Hyperparameter} \\
\cline{2-11} 
	 & $I$ & $J$ & $K$  & $\alpha$ & $\beta$ & $\gamma$ & $\delta$ & $\eta$ & $\rho$  & $\omega$ \\
\hline
Multi & 4 & 8 & 5  & 1 & 1 & 10 & 1 & 10 & 1 & 5e-5 \\
\hline
Multi$^\dag$ \hspace*{-7pt} & 4 & 8 & 5  & 1 & 1 & 0 & 0 & 0 & 0 & 5e-5 \\
\hline
Single & 1 & 32 & -  & 1 & 1 & 10 & 1 & 10 & 1 & 5e-5 \\
\hline
Single$^\dag$ \hspace*{-7pt} & 1 & 32 & -  & 1 & 1 & 0 & 0 & 0 & 0 & 5e-5 \\
\hline
Vanilla & 1 & 32 & -  & 1 & 1 & 10 & 1 & 10 & 1 & 5e-5 \\
\hline
Vanilla$^\dag$ \hspace*{-7pt} & 1 & 32 & - & 1 & 1 & 0 & 0 & 0 & 0 & 5e-5 \\
\hline
\end{tabular}
\vspace*{10pt}
\caption{Hyperparameter settings of neural network models, where all models use the same learning rate 0.1 and the same number of iterations 2e+4.}
\label{tab:hyperparam}
\end{table}

\begin{table}[t]
\centering
\begin{tabular}{c|cc|cc}
\hline
\multirow{2}{*}{Model}	& \multicolumn{2}{c|}{Training} & \multicolumn{2}{c}{Test} \\
\cline{2-5}	
				&  Mean &  STD &  Mean &  STD\\ 
\hline
Multi		&        1.74 &        0.50 &       3.34 &       2.18 \\
Multi$^\dag$ 		&        1.76 &        0.50 &       3.35 &       2.17\\
Single 		&        2.15 &        0.67 &       3.60 &       2.12 \\
Single$^\dag$ 	&        1.82 &        0.52 &       3.38 &       2.16  \\
Vanilla 	&        3.21 &        0.98 &       4.46 &       2.07 \\
Vanilla$^\dag$ 	&        2.87 &        0.80 &       4.18 &       2.04  \\
SSVI 		&        2.59 &        0.85 &       3.73 &       2.18  \\
\hline
\end{tabular}

\vspace*{10pt}

\caption{Mean and standard deviation (STD) of the MAPEs for the predicted implied volatilities from the examined models.}
\label{tab:mape_overall_iv}

\vspace*{10pt}

\begin{tabular}{c|cc|cc}
\hline
\multirow{2}{*}{Model}	& \multicolumn{2}{c|}{Training} & \multicolumn{2}{c}{Test} \\
\cline{2-5}	
				&  Mean &  STD &  Mean &  STD\\
\hline 
Multi		    &       5.97 &        1.86 &      10.64 &       6.72 \\
Multi$^\dag$ 	&       6.03 &        1.86 &      10.67 &       6.70 \\
Single 		    &       7.38 &        2.57 &      11.64 &       6.68 \\
Single$^\dag$ 	&       6.20 &        1.91 &      10.77 &       6.67 \\
Vanilla 		&       11.31 &        3.57 &      14.61 &       6.42 \\
Vanilla$^\dag$ 	&       10.53 &        3.34 &      14.17 &       6.60 \\
SSVI 			&       8.71 &        2.72 &      12.74 &       6.74 \\
\hline
\end{tabular}
\vspace*{10pt}
\caption{Mean and standard deviation (STD) of the MAPEs for the option prices calculated using the predicted implied volatilities from the examined models.}
\label{tab:mape_overall_op}
\end{table}

\subsection{Experimental Settings} 
\label{sec:settings}

Table~\ref{tab:models} summarises the examined models in the experiments. In addition to our proposed model (simply denoted by Multi), we also deploy the SSVI model, and other neural network models. For the former, we aim to investigate if our proposed model can achieve a better prediction performance than the state-of-the-art method from mathematical finance. For the latter, we want to see if the designed architecture (i.e., the ensemble of multiple single networks) can improve the prediction capability. To this end, the benchmarked neural networks include: 1) a single network (denoted by Single) where there is no weighting network ; and 2) a simplest neural network (denoted by Vanilla) which has a single hidden layer, only uses the sigmoid activation function and the model's output is censored to be non-negative. Also, to further justify the importance of embedding financial conditions, all the neural network models are trained under a setting where $\ell_1, \ell_2, \ell_3, \ell_4$ are removed from the total loss function $\ell$. To simplify the discussion, these neural network models are denoted with a superscript $\dagger$.   

Table~\ref{tab:hyperparam} presents our hyperparameter settings of the examined neural networks. To avoid the size effect on model performance, the compared neural networks with the same architecture design are specified with the same model size and with the same hyperparameter values. As also mentioned earlier in Section~\ref{sec:model}, embedding $\ell_1, \ell_2, \ell_3, \ell_4$ requires synthetic data. In our neural network training, the ratio of real market data and synthetic data is 1/6. The log forward moneyness $m$ is sampled in $[-6,-3] \cup [3,6]$ for the asymptotic condition and in $[-3,3]$ for other conditions; and the time to maturity $\tau$ is sampled in $[0.002,3]$. These values are set based on the observations in Figure~\ref{fig:data_summary}. Neural network models are trained using TensorFlow and we use Adam~\cite{Kingma15adam} for stochastic optimisation. We train the models using all option quotes from the previous trading day and test the models in the next day. We move the window of training and test split across all the trading days in the option data.

\subsection{Results and Analysis}
\label{sec:results}

\begin{figure}[t]
\centering
\includegraphics[width=1\linewidth]{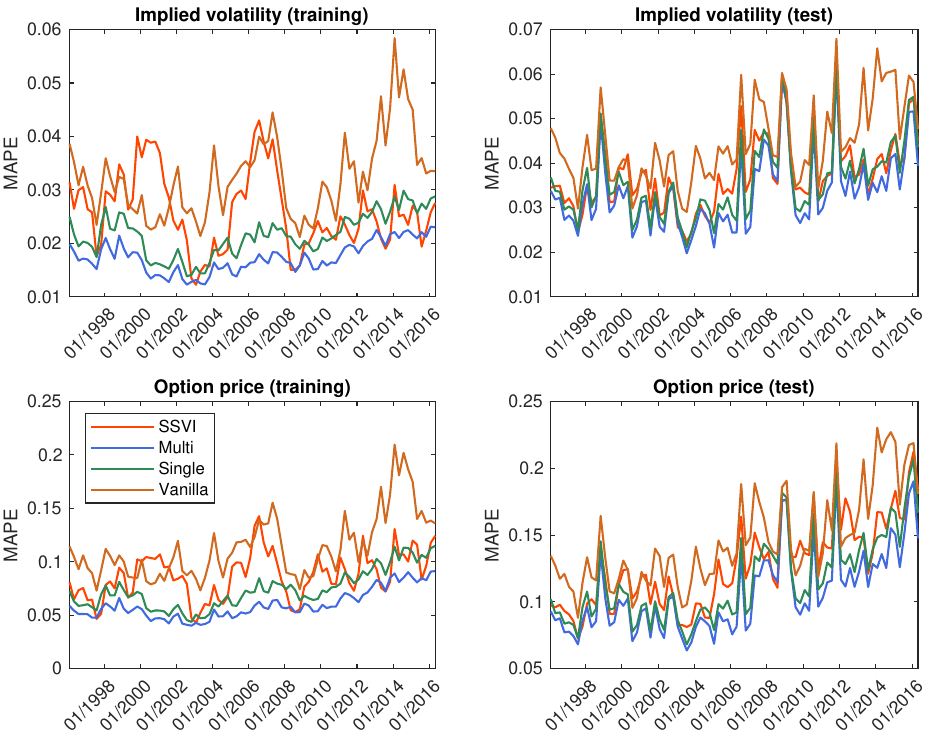}
\caption{Time series plot of the MAPEs for the SSVI model and the neural network models trained with full settings.}
\label{fig:season}
\end{figure}

The summary statistics of the mean average percentage errors (MAPEs) of the predicted implied volatilities for the examined models in all trading days are presented in Table~\ref{tab:mape_overall_iv}. To further investigate the effects or differences that the predicted implied volatilities can trigger in option pricing, we use them to compute the corresponding option prices and then report the summary statistics of the MAPEs of the option prices in Table~\ref{tab:mape_overall_op}. In both tables, the widely used SSVI model from mathematical finance underperforms the Multi, Multi$^\dagger$, Single and Single$^\dagger$ models significantly but it still outperforms the simple neural networks like the Vanilla and Vanilla$^\dag$ models. Compared to conventional mathematical models in finance, data-driven deep learning models have shown great predictive capabilities but they should be with a proper architecture design and hyperparameters settings. Our study has been successfully validated with the results because the proposed Multi model is the best-performing prediction model in both training and test data for implied volatility and option price, respectively. Figure~\ref{fig:season} also compares the time series plots of the MAPEs of the SSVI model and the deep neural networks trained with the full settings on each sampled trading day, showing our proposed Multi model has the most stable outputs over time. 

\begin{figure}[t]
\centering
\includegraphics[width=0.9\linewidth]{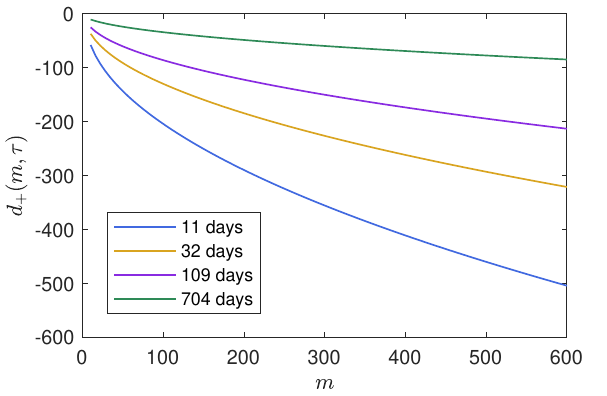}
\caption{Limiting behaviour of the converted forward contracts with 11, 32, 109, and 704 days duration.}
\label{fig:limit}

\vspace*{20pt}

\includegraphics[width=1\linewidth]{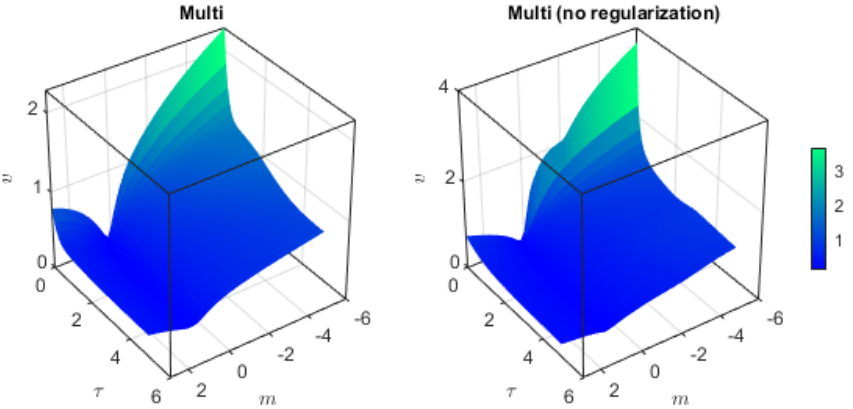}
\caption{Implied volatility surface on 11/01/2016 predicted by: (a) the multi model; and (b) the multi model without regularization.}
\label{fig:iv_surface}
\end{figure}

One interesting finding from Tables~\ref{tab:mape_overall_iv}-\ref{tab:mape_overall_op} is that excluding some financial conditions in training neural networks will not significantly decrease the models' prediction performance because the Multi$^\dagger$, Single$^\dagger$ and Vanilla$^\dag$ models have a comparable performance with their counterparts which were trained with the full settings (i.e., the Multi, Single and Vanilla models). However, as discussed previously, incorporating the prior financial domain knowledge is mainly to ensure the model is consistent with the existing financial theories and assumptions rather than the model's prediction performance. In Table~\ref{tab:conditions}, we check if the monotonicity, boundary, absence of butterfly arbitrage and asymptotic slope conditions in Theorem~\ref{thm:conditions} are empirically satisfied in the test data. Since we have discussed in Section~\ref{sec:model} that the positivity and twice differentiation conditions are met in our neural network architecture design and the limiting behaviour condition can be proven theoretically, these three conditions are not checked by Table~\ref{tab:conditions}. It is not difficult to observe that the violation percentage of the test samples for the examined conditions of the Multi$^\dagger$, Single$^\dagger$ and Vanilla$^\dag$ models are much higher than the corresponding Multi, Single and Vanilla models, showing the importance and necessity of embedding $\ell_1, \ell_2, \ell_3, \ell_4$ loss functions. For illustration purpose only, Figure~\ref{fig:limit} demonstrates the limiting behaviour of the converted underlying forward contracts with 11, 32, 109, and 704 days duration, verifies the limiting behaviour condition. 

Further, to get a sense of what an implied volatility surface looks like for our readers, Figure~\ref{fig:iv_surface} shows the surfaces resulting from the Multi model and the Multi model trained without the regularization term for 11/01/2016. It is clear that the volatility smiles in the former are much smoother than those of the latter, which also verifies the effectiveness of regularization. 

\begin{table*}[t]
\centering
\begin{tabular}{c|c|c|c|c|c}
\hline 
Model	  & Monotonicity & Absence of butterfly arbitrage  & Left boundary		& Right boundary		& Asymptotic slope\\ 
\hline
Multi	  & 0.00\%	& 7.02e-6\%	 &  0.00\% & 0.00\%	& 0.00\% \\
\hline
Multi$^\dag$ 	&   1.28\%	& 4.87\%  & 0.00\%	&14.06\%	 & 0.00\% \\
\hline
Single	&   0.00\% 	& 5.56e-3 \% &  0.00\% 	& 0.05\%	&  0.00\%  \\
\hline
Single$^\dag$ 	& 0.00\%	& 14.88\% & 0.95\% & 5.16\%	 & 0.00\% \\
\hline
Vanilla	& 3.75e-3\%  & 1.53e-2\% & 4.07e-3\%  & 	0.00\%	&	0.00\% \\
\hline
Vanilla$^\dag$ 	& 5.32\% & 5.72\%	 & 14.63\%	& 0.00\%	 & 0.54\% \\
\hline
\end{tabular}
\vspace*{10pt}
\caption{Percentages of the test set samples which do not satisfy conditions 3,4,6,7,8 of Theorem~\ref{thm:conditions}.}
\label{tab:conditions}
\end{table*}

\section{Conclusion}
\label{sec:conclusion}

In this paper, we developed a novel neural network to predict implied volatility surfaces. Unlike many previous studies where the machine learning algorithms are mainly used as the \lq\lq{}black box\rq\rq{} in finance, our model is tailored to the unique characteristics of implied volatility surface. To the best of our knowledge, this is one of the very first studies which discuss a methodological framework that integrates the data-driven machine learning algorithms (particularly neural networks) with the related financial theories and empirical evidence. The proposed model framework can be easily extended and applied to solve other similar computational problems in finance and business analytics such as inventory pricing and revenue management. 

In addition to the methodological contribution, we validated the proposed model empirically with the option data on the S\&P 500 index. Compared with the existing studies, our experimental settings are more challenging because the used option data is over 20 years and the options with the short time to maturity are examined. Therefore, our model needs to be more robust in order to produce convincing results. As presented in the experiments section, our model outperforms the widely used SSVI model from mathematical finance and other benchmarked neural networks. More importantly, the conventional financial conditions and empirical evidence are met empirically, which resolve the bottleneck of data-driven machine learning applications in finance.


\begin{acks}
This research has been conducted with the support of:  1) the Financial Innovation Center of the Southwestern University of Finance and Economics; and 2) the Key Laboratory of Financial Intelligence and Financial Engineering of Sichuan Province. 3) the UK Economic and Social Research Council through the Impact Acceleration Accounts Business Booster Funding to the University of Glasgow. The first author also acknowledges the Imperial College Business School with the support of the high performance computing equipment for experiments during his PhD study~\cite{Zheng_2018}. The authors would also like to thank anonymous reviewers for their helpful comments on earlier drafts of the manuscript.
\end{acks}

\bibliographystyle{acm}
\balance
\bibliography{nnivs_short}

\end{document}